\documentclass[letterpaper,10pt,conference]{IEEEtran}

\IEEEoverridecommandlockouts

\usepackage{lipsum}
\usepackage{graphics,graphicx}
\usepackage{fancyhdr,lipsum}

\makeatletter
 \let\old@ps@headings\ps@headings
 \let\old@ps@IEEEtitlepagestyle\ps@IEEEtitlepagestyle
 \def\confheader#1{%
 \def\ps@headings{%
 \old@ps@headings%
 \def\@oddhead{\strut\hfill#1\hfill\strut}%
 \def\@evenhead{\strut\hfill#1\hfill\strut}%
 }%
 \def\ps@IEEEtitlepagestyle{%
 \old@ps@IEEEtitlepagestyle%
 \def\@oddhead{\strut\hfill#1\hfill\strut}%
 \def\@evenhead{\strut\hfill#1\hfill\strut}%
 }%
 \ps@headings%
 }
 \makeatother

\confheader{%
2023 IEEE World AI IoT Congress (AIIoT)
 }

 \usepackage[pscoord]{eso-pic}
\newcommand{\placetextbox}[3]{
 \setbox0=\hbox{#3}
 \AddToShipoutPictureFG*{ \put(\LenToUnit{#1\paperwidth},\LenToUnit{#2\paperheight}){\vtop{{\null}\makebox[0pt][c]{#3}}}
 }
 }
 \placetextbox{.23}{0.055}{\small{979-8-3503-3761-7/23/\$31.00~\copyright 2023 IEEE}}

\begin{document}

\title{Anomaly Detection Techniques in Smart Grid Systems: A Review}

\author{
\IEEEauthorblockN{Shampa Banik}%
\IEEEauthorblockA{
\textit{IEEE Student Member}\\
\textit{Dept. of Computer Science} \\
\textit{Tennessee Tech University}\\
Tennessee, USA \\
sbanik42@tntech.edu}
   
   \and
   
\IEEEauthorblockN{Sohag Kumar Saha}%
\IEEEauthorblockA{
\textit{IEEE Student Member} \\
\textit{Dept. of ECE } \\
\textit{Tennessee Tech University}\\
Tennessee, USA \\
ssaha42@tntech.edu}

\and

\IEEEauthorblockN{Trapa Banik}%
\IEEEauthorblockA{\textit{IEEE Student Member} \\
\textit{Dept. of ECE} \\
\textit{Tennessee Tech University}\\
Tennessee, USA \\tbanik42@tntech.edu}

\and

\IEEEauthorblockN{S M Mostaq Hossain}%
 \IEEEauthorblockA{
 \textit{IEEE Student Member} \\
 \textit{Dept. of Computer Science} \\
\textit{Tennessee Tech University}\\
Tennessee, USA \\shossain42@tntech.edu}

}

\maketitle


\pagestyle{fancy}
\fancyfoot[C]{\thepage}

\maketitle

\begin{abstract}

Smart grid data can be evaluated for anomaly detection in numerous fields, including cyber-security, fault detection, electricity theft, etc. The strange anomalous behaviors may have been caused by various reasons, including peculiar consumption patterns of the consumers, malfunctioning grid infrastructures, outages, external cyber-attacks, or energy fraud. Recently, anomaly detection of the smart grid has attracted a large amount of interest from researchers, and it is widely applied in a number of high-impact fields. One of the most significant challenges within the smart grid is the implementation of efficient anomaly detection for multiple forms of aberrant behaviors. In this paper, we provide a scoping review of research from the recent advancements in anomaly detection in the context of smart grids. We categorize our study from numerous aspects for deep understanding and inspection of the research challenges so far. Finally, after analyzing the gap in the reviewed paper, the direction for future research on anomaly detection in smart-grid systems has been provided briefly.

\end{abstract}

\section{INTRODUCTION}

\IEEEPARstart{S}{mart Grid} paradigm integrates smart technology in areas such as information and communication (ICT), control, sensing \& measurement, energy storage, automation, distributed generation, renewable generation \& integration that conventional grids do not possess\cite{gopstein2021nist}. Fig. 1 shows the multi-disciplinary fields of the smart grid developed by the National Institute of Standards and Technology (NIST) and Smart Grid Interoperability Panel (SGIP). Wide deployment of the ICT infrastructure acts as the backbone of these intelligent technologies to enable a bidirectional data flow between the utility and the users. With the implementation of advanced metering infrastructure (AMI) technologies, two-way electricity and information communication would be enabled, empowering the SG with near-real-time monitoring/display of consumer energy consumption data, decision-making, dynamic tariff negotiations, and load control for a variety of applications. The widespread adoption of different types of electric appliances and the rapid development in the integration of renewable energy generation has created new challenges for SG systems. The word "anomaly" is used throughout this paper to refer to any abnormal or strange behavior or event. These anomalies could originate from a number of sources, including customers' unusual consumption patterns, infrastructure failures, power outages, malicious cyber-attacks, or energy theft. 
The power outage, transmission line outage, unusual power consumption, and momentary and sustained outages are all examples of the kind of anomalies that can be considered to fall under the umbrella term "anomaly" \cite{lipvcak2019big}. 
Improving grid networks' operational effectiveness and dependability depends on developing efficient anomaly detection algorithms for identifying these anomalous behaviors. Anomaly detection, also known as outlier detection, is a technique used to identify outliers—extreme cases in a dataset that stand out from the norm.  Data collected from networks, IoT devices, SCADA systems, smart grids, or the logs of any machine can now be categorized using Artificial Intelligence (AI) technologies into either a binary classification of normal or abnormal behaviors or a multi-class attack classification of normal or varying types of abnormal behaviors associated with specific attacks\cite{elmrabit2020evaluation}. 
Numerous suggestions in the literature attempt to solve the anomaly detection problem.
Besides the wide ranges of machine learning based approach \cite{panthi2020anomaly, liu2016regression, himeur2021artificial}, time series based \cite{zhang2021time, hyndman2015large}, the grid load variations \cite{wei2020glad}, the IoT premises \cite{gaddam2019anomaly,erhan2021smart, marino2019cyber}, modbus network vulnerabilities attack  \cite{shampa2023new} and anomaly based on smart meter data \cite{liu2018scalable, jaiswal2020distributed, yuan2015distributed, jaiswal2021anomaly, yen2019effect} are other means of analyze anomalies in smart grid system. In addition to those another approach such as digital twins \cite{digitaltwinmostaq2023} in smart grids can detect anomalies by comparing real-time data against established baselines, analyzing sensor data, monitoring grid performance, identifying cyber-security threats, correlating events, and utilizing predictive analytics.

\begin{figure}[h!]
\centering
\includegraphics[width=1\columnwidth]{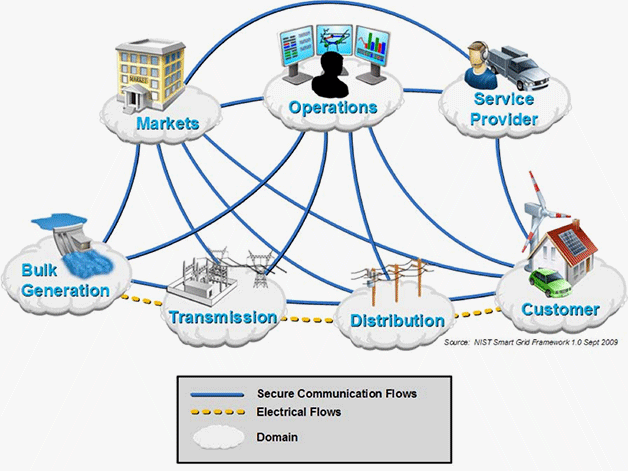}
\caption{The multi-disciplinary fields of smart grid \cite{gopstein2021nist}}
\label{Fig. 1}
\end{figure}

The contribution of this paper is as follows:
1) to investigate the different recent pertinent research works on anomaly detection techniques 2) to identify potential gaps research gaps that has not been explored 3) to recommend one potential approach as a detection technique of anomalies in smart grid. Initially, we present the approach of anomaly detection study by categorizing. Then besides discussion, critical analysis of the limitations and strengths of recent pertinent literature have been analyzed. Finally, we recommend the best-fit approach for facing the anomaly detection challenge as a future research direction. 

In particular, the rest of this paper is structured as follows:
Section II discusses the research background of the literature and provides the categories of research problems and solution methodologies of the related papers, along with an analysis of related papers by investigating their main characteristics. Section III outlined the prevalent recent anomaly detection mechanism in SG and future research direction. Section IV provides the new technique of anomaly detection with its challenges and contributions. Finally, section V presents the concluding summary of this study.

\section{RESEARCH OVERVIEW}

To delve deeper into anomaly detection in the SG, we intended to categorize our findings in this paper from different research perspectives. Fig. 2 depicts the schematic diagram of our study for anomaly detection in SG on the basis of detection level, anomaly types, and computing methodology.
The first category deals with the level of detection which is pervasive nowadays, such as cyber-attacks, various faults, theft attacks, and cyber-physical systems. Besides these, we also tried to spotlight the types of anomalies associated with power supply rate, customer behavior, load balancing, and power outages. The other category is intended for computing which is incorporated with the respective methodologies in detecting anomalies, for example, the machine learning approach, cloud computing, hybrid computing, and edge computing.

The third category is devoted to the pertinent findings on anomaly detection that have evolved with the reasons behind it, such as cyber-physical threats that are too common in today’s perspective. Again, when it comes to fault and energy theft detection, which is another critical level of detection for an anomalous SG, These types of attacks make the SG system vulnerable to fluctuations in power supply, which can lead to power outages at times. For example, the power grid of Ukraine was hacked in 2015, which resulted in power outages \cite{case2016analysis}. Most importantly, because the hacker or attacker is currently attempting to manipulate customer meter data, unusual customer behavior is another anomaly indicator. To move forward, we have gone through a substantial number of studies in this field that have necessarily come up with some novel ideas with the aim of solving the problem to a certain extent. The evaluation of the devised strategies is based on deep analysis rather than via pilot projects and via analytical and simulation means. Adding contextual information and including an event-type instance in relation to other event-type instances proved to be a more powerful mechanism. Several papers have investigated the IoT and SG security issues. Some remarkable cases are listed in cyber-physical systems or hybrid computing. We also present a comprehensive study on cloud computing and machine learning in SG.

\begin{figure*}[h!]
\centering
\includegraphics[scale=0.9]{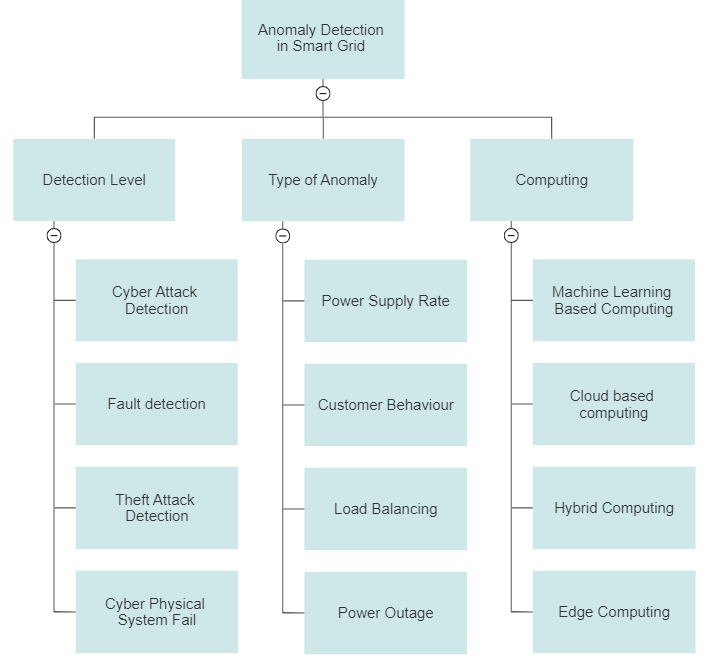}
\centering\caption{Categories of the anomaly detection study in smart grid based on different aspects: (i) detection levels, (ii) type of anomaly, (iii) computing platforms}
\label{Fig. 2}
\end{figure*}

\subsection{Literature Background}

In the context of SG security, a plethora of analyses has been done on anomaly detection in SG which is the most crucial reason behind the anomaly.

\textbf{Cyber-Attack Based:} The FDI attack was identified in \cite{ibrahem2021detecting} using a processed benign data set and a collection of attacks. The suggested system employed deep learning to train datasets and detect assaults. In order to detect multiple stealthy cyber attacks on DER networks, Moataz et al. \cite{abdelkhalek2022ml}recommended a supervised machine learning-based anomaly detection (ML-ADS) algorithm, and a testbed has been implemented with results showing high detection accuracy (98.4\%), very low detection latency (5microseconds), and high precision.

\textbf{SM Data Based:} From the SM data perspective there are a recent substantial amount of research work has been found. The authors in \cite{moghaddass2017hierarchical} implemented a real-time anomaly detection system that is based on data gathered from SMs installed at customers' homes. The system is basically a parameter estimation method for model training and then discussed how the trained structure can be used for anomaly detection.
For an AMI infrastructure, Li and his team \cite{li2022look} proposed a privacy-preserving anomaly attack detection system that uses a look-up-table (LUT) based fully homomorphic encryption (FHE) integrated with private information retrieval (PIR), with the framework being implemented on a Raspberry PI standing in for a SM. Based on the experimental results, it was determined that the proposed approach can identify the injection of erroneous power consumption in 11–17 seconds of execution time, with the precision of detection determining the exact time. In \cite{liu2016smart}, Golab and his peers are more concerned with what can be done with the SG data than with how efficiently it can be done. For this, they have created a performance benchmark for typical SM analytics operations and described an approach to generate huge realistic datasets from a tiny seed of genuine SM data.

\textbf{Power Consumption Based:}
To improve transfer learning's ability to detect spinning power consumption anomalies, the authors proposed a CDAN model \cite{xu2021anomaly} a cluster-based adaptation layer is added between source and target network feature layers. Another analysis has been found to alert utility companies to suspicious consumption behavior that could be further investigated with on-site inspections or other methods. These have been possible by applying two types of Machine Learning (ML) algorithms for anomaly detection for unlabeled data\cite{panthi2020anomaly}: Spectral Residual-Convolutional Neural Network (SR-CNN) and an anomaly-trained model on the basis of martingales for determining variations in time-series data streams. A framework is developed to identify electricity theft using real-time SG data in \cite{tehrani2022online}. Building managers may benefit from Janetzko et al. \cite{janetzko2014anomaly} analytical and visualization method, which they used for the anomalous time-series power consumption data pattern for a residential building. Power consumption anomalies are identified by either a clustering-based approach or a time-weighted prediction, and both methods are contrasted with a similarity-based anomaly computation/detection process. Abdel and his group \cite{abdel2022privacy} pondered this issue by incorporating a  semisupervised DL (SSDL) framework, termed as federated semi-supervised class-rebalanced (Fed-SCR) that is particularly employed to identify the anomaly that occurred in power data due to fog in an industrial SG.  In the paper authored by Liu and Nielsen \cite{liu2016regression}, a SG consumption pattern-based supervised learning and statistical anomaly detection approach is proposed that detects aberrant energy usage from one's consumption history where the lambda architecture is deployed for model updating and real-time anomaly detection.

\textbf{Machine Learning Based:}
The research suggested an anomaly detection system (ADS) based on supervised machine learning (ML) to identify covert stealthy IT and OT-based attacks on DER communication in \cite{abdelkhalek2022ml}.
A hybrid strategy to identify fraud combines data and network-oriented techniques. The authors created a SG electricity theft detection framework. Density-based abnormal detection, described by Zheng et al. \cite{zheng2017electricity}, calculates the abnormal degree of user profiles based on their distance matrix and beats other unsupervised approaches including k-means clustering, GMM clustering, and DBSCAN.
In another journal \cite{siniosoglou2021unified}, the authors presented an Intrusion Detection System (IDS) called MENSA (Anomaly Detection and Classification) that adopts a novel Autoencoder-Generative Adversarial Network (GAN) architecture for (a) abnormalities in operation detection and (b) cyber threat classification using Modbus/TCP and DNP3. With the adversarial loss and the reconstruction difference in mind, MENSA incorporates the aforementioned Deep Neural Networks (DNNs) in a single design. Rituka and her co-workers \cite{jaiswal2021anomaly} implemented Generalize Additive Regression Model using Prophet, a Facebook-developed library  that uses confidence intervals to detect abnormalities to avert power outages and preserve grid balance. Another novel approach postulated by Himeur et al. \cite{himeur2020novel}for detecting energy consumption anomalies is based on extracting micro-moment information using a rule-based model and a deep neural network (DNN) architecture for fast anomaly identification and classification. Machine learning was used by Reddy and co-authors \cite{shabad2021anomaly} by adopting an Isolation Forest based model that makes use of historical data to make predictions about future grid parameters and using principal component analysis to optimize feature selection. 
A recurrent generative adversarial network (R-GAN) is presented by Fekri et al. \cite{fekri2019generating}to learn from actual data and generate realistic energy usage statistics. Wasserste in GANs (WGANs) and Metropolis-Hastings GAN (MH-GAN) methods were employed to address convergence instability and boost the quality of produced data, respectively whereas new features were created using ARIMA and the Fourier Transform to enhance the quality of the produced data. A system based on prediction-based detection was proposed by Liu et al. \cite{liu2018scalable}and compared to three baseline approaches for processing massive amounts of data on a cluster using a unique lambda architecture. Sohrab and colleagues \cite{mokhtari2021machine} suggested a supervised machine learning model called a measurement intrusion detection system (MIDS) to distinguish between typical and unusual Industrial Control System (ICS) operations and assess the MIDS's efficacy.

\textbf{Cloud Computing Based:}
Fog computing is another efficacious approach for real-time big data as compared to cloud computing while dealing with a humongous number of sensors. Jaiswal's team \cite{jaiswal2020distributed} proposed a hierarchical Distributed Fog Computing architecture for rolling out anomaly detection models trained by machine learning to extract information from residential SM sensor readings. Anomaly detection is split into two stages: (1) model training using an ensemble of prediction-based models, and (2) the removal of spurious anomalies using the Ensemble approach of majority voting. The purpose of the study  executed by Ferrari and his fellows \cite{ferrari2019performance} is to provide an experimental approach to contrasting edge-cloud and full-cloud architectures for deep learning-based anomaly detection which emphasizes data transmission latency and Cloud bandwidth measurements. 

\textbf{IoT-Based:}
For an IoT-based setup, Yuan et al. \cite{yuan2015distributed} proposed a deep learning-oriented distributed anomaly detection method in which massive monitoring data is automatically extracted using a Stacked sparse autoencoder and then categorized using Softmax to detect the anomaly and issue a warning to the master device.
In the study\cite{tehrani2022online}, the idea has been shifted from a single event being classified as an anomaly to the idea of a collective anomaly, or a group of events that may be abnormal based on their patterns of occurrence.

\textbf{Time Series Based:}
Another significant issue known as time series edge anomaly related to SG data transmission through cloud platforms is addressed by Xu et al. \cite{xu2021graph}. A graph convolution neural networks (GCNs) based edge-cloud collaborative anomaly detection model was proposed to rectify the problem where the time series data is transformed into graphical data by time series directed horizontal visibility graph algorithm and a model segmentation approach is developed to create efficient model inference.

\textbf{Simulation-Testbed Based:}
The approach in \cite{karimipour2019intelligent} is to develop a testbed that mimics consumer behavior and incorporates hardware into the loop to detect the anomaly in SG. With the help of this testbed, the test, and assessment of some potential attacks they tried to demonstrate how components and customer behavior interact with one another and with the software that runs on them. 

Yen et al. \cite{yen2019effect} developed a  MATLAB/Simulink-based anomaly detection method to detect voltage normalcy at diverse locations of a smart power system  to detect voltage abnormalities in the power system utilizing huge SM data acquired at multiple data collection frequencies.

To model the power-generating equipment and use the attack dataset, a hardware-in-the-loop (HIL) testbed has been built. Zhang et al. \cite{zhang2023improve} also proposed a Neural Network architecture called MODLSTM to improve ICS security, particularly for DoS attacks. They implemented the design on industrial and public datasets and found that the models performed better than previous work (accuracy increased by 0.71\% and 0.07\%, respectively).

Wasim and Haroon \cite{khan2022efficient}hypothesized that their proposed novel framework captures cross-modality interactions between network structure and node attributes as well as includes the potential data distribution pattern by dual variational autoencoder with the ability to  detect unlabeled anomalies by Gaussian Mixture Model.

The discussion of the aforementioned numerous technologies of anomaly detection is further analyzed in Table 1. By summarizing the main/key research concept, strength points, and limitations of each of the research papers/articles, the following important information from selected studies is tabulated.

\begin{table*}[hbt!]
\caption{Summary of the paper reference based on the Main Concept, Feature, and Limitation}

\begin{tabular}{|l|l|l|l|l|}
\hline
\textbf{SL\#} & \textbf{Implemented technique(s)} & \textbf{Summary of Core Idea} & \textbf{Point of Strength} & \textbf{Limitation} \\


\hline

\multicolumn{1}{|p{0.4em}}{1} &    
\multicolumn{1}{|p{13em}|}{Deep Learning \cite{tang2022research} } &
\multicolumn{1}{p{15em}|}{Net-metering system issues are examined here. The system's datasets are trained, and threats are detected using deep learning.} & \multicolumn{1}{p{14.5em}|}{The multidata-source detector may reliably identify false-reading attacks, according to two trials.} & \multicolumn{1}{p{14em}|}{Consumers' house conditions, appliances, and sleeping patterns revealed through net meter readings.} \\

\hline

\multicolumn{1}{|p{0.4em}}{2} &
\multicolumn{1}{|p{13em}|}{Transfer Learning\cite{xu2021anomaly}} & 
\multicolumn{1}{p{15em}|}{A cluster-based adaptation layer is introduced between the source and target networks' feature layers in the Cluster-based Deep Adaptation Network (CDAN) paradigm.} & \multicolumn{1}{p{14.5em}|}{Experiments demonstrate that the suggested approach can identify spinning power consumption abnormalities more accurately than existing methods.} & \multicolumn{1}{p{14em}|}{The anomaly detection method was not optimized for operational data in this study.} \\

\hline

\multicolumn{1}{|p{0.4em}}{3} &
\multicolumn{1}{|p{13em}|}{Machine Learning\cite{abdelkhalek2022ml}} &
\multicolumn{1}{p{15em}|}{This study proposed a supervised machine learning-based anomaly detection system (ADS) to detect stealthy IT and OT-based attacks on DER communication.} & \multicolumn{1}{p{15em}|}{The ANN-based ADS demonstrated 98.4\% detection accuracy, strong recall, and precision in a 2-tier IADS end-to-end DER architecture.} & \multicolumn{1}{p{14em}|}{For cyber situation awareness across an entire system, there are no state-of-the-art ML-based event correlation methods.} \\

\hline

\multicolumn{1}{|p{0.4em}}{4} &
\multicolumn{1}{|p{13em}|}{Hybrid Approach\cite{tehrani2022online}} & \multicolumn{1}{p{15em}|}{ A hybrid strategy for identifying fraudulent conduct is provided, combining data-oriented and network-oriented techniques.} & \multicolumn{1}{p{14.5em}|}{This framework pre-processes data, does feature engineering, discovers consumption patterns, classifies consumption, and checks user data for fraud. It lowers false alerts.} & \multicolumn{1}{p{14em}|}{Inserted assaults are 73.8–94 percent accurate, excellent in detecting things but require an extensive amount of data samples, don't function distributively, and are hard to keep private.} \\

\hline

\multicolumn{1}{|p{0.4em}}{5} &
\multicolumn{1}{|p{13em}|}{Big Data Platform\cite{lipvcak2019big}} & \multicolumn{1}{p{15em}|}{The platform is built on an ingestion layer with options for data compression, Apache Flink as part of the speed layer, and HDFS/KairosDB as data storage layers.} & \multicolumn{1}{p{14.5em}|}{ Flink was chosen for inclusion in the platform's speed layer because it offered the best performance for stream processing and satisfied the needs for anomaly detection in power consumption datasets.} & \multicolumn{1}{p{14em}|}{Comparison is absent for several anomaly detection algorithms to help better identify clusters of customers based on smart metering data traces.} \\
       
\hline

\multicolumn{1}{|p{0.4em}}{6} &
\multicolumn{1}{|p{13em}|} {Unsupervised Machine Learning with IREST sensor \& ISAAC Testbed\cite{marino2019cyber} } & \multicolumn{1}{p{15em}|}{The idea is based on a cyber-physical sensor called IREST (ICS Resilient Security Technology)which underwent a range of cyber-physical testing on the Idaho CPS SCADA Cybersecurity (ISAAC) testbed.} & \multicolumn{1}{p{14.5em}|}{Unsupervised machine learning (ML) finds CPS cyber and physical anomalies. The IREST cyber-sensor and ISAAC testbed provide a powerful and scalable framework for cyber-physical security research.} & \multicolumn{1}{p{14em}|}{Lacking for several local IREST sensor analytics in large-scale distributed HIL simulations.}\\

\hline

\multicolumn{1}{|p{0.4em}}{7} &
\multicolumn{1}{|p{13em}|}{Unsupervised Learning \cite{karimipour2019intelligent}} & \multicolumn{1}{p{15em}|}{The objective is to locate bizarre objects without human assistance. Unsupervised anomaly detection for SGs may identify FDI assaults using prior measurement data.} & \multicolumn{1}{p{14.5em}|}{Simulation results on IEEE 39, 118, and 2848 bus systems demonstrate the method's versatility and  99\% accuracy, 98\% true positive, and ±2\% false positive.} & \multicolumn{1}{p{14em}|}{The attacks were detected using various IEEE bus systems using unsupervised learning. It can also be tested in various linear and non-linear power systems to make it more accurate and precise.} \\
    
\hline

 \multicolumn{1}{|p{0.4em}}{8} &     
\multicolumn{1}{|p{13em}|}{ Blockchain Technology+ Big Data + Machine Learning \cite{li2020blockchain}} & \multicolumn{1}{p{14.5em}|}{A new framework using sensor processing, SM readings, machine learning, and blockchain is introduced before the system and monitoring models.} & \multicolumn{1}{p{15em}|}{Two similar methods are investigated in IWSN's big data and machine learning-based power consumption anomaly detection framework.} & \multicolumn{1}{p{14em}|}{Meteorological data and industrial operator operating trends may enhance data granularity.} \\
        
\hline

\multicolumn{1}{|p{0.4em}}{9} &
\multicolumn{1}{|p{13em}|}{Deep Autoencoder+ Long Short-Term Memory (LSTM)\cite{takiddin2022deep}} & \multicolumn{1}{p{14.5em}|}{Deep (stacked) autoencoders with a sequence-to-sequence (seq2seq) structure based on long-short-term memory are proposed (LSTM). Simple, variational, and attention-based autoencoders were evaluated (AEA).} & \multicolumn{1}{p{15em}|}{The suggested solution used benign power consumption profiles with numerous deep autoencoders to address the lack of harmful profiles. Seq2seq structures outperform fully linked ones for detection.} & \multicolumn{1}{p{14em}|} {As the malicious datasets are not readily available, therefore the pre-processed benign datasets are used only.} \\

\hline

\multicolumn{1}{|p{0.4em}}{10} &
\multicolumn{1}{|p{13em}|} {Homomorphic Encryption (HE)+Custom fault detection framework using AMIs \cite{ishimaki2020towards}} & \multicolumn{1}{p{15em}|}{AMIs use homomorphic encryption (HE) to identify anomalies in encrypted SM data. Customers' fine-grained SM measurements are never disclosed.} & \multicolumn{1}{p{14.5em}|}{Ratio calculation data encoding shrinks HE ciphertext. With the HE parameter set to spot irregularities in SM information, our simple, efficient user-side encryption solution for low-power devices is 40x faster than a naïve one.} & \multicolumn{1}{p{14em}|}{The quantification of the fine-grained power consumption information leakage from the HM-AM ratio will be a future task to develop.} \\

\hline

\multicolumn{1}{|p{0.4em}}{11} &
\multicolumn{1}{|p{13em}|} {Linear programming (LP) based detection framework using AMIs \cite{yip2018anomaly}} & \multicolumn{1}{p{15em}|}{
An anomaly detection framework that can effectively identify energy theft attacks against AMI is now crucial for mitigating financial losses ncurred due to uncontrolled non-technical losses (NTLs).} & \multicolumn{1}{p{14.5em}|}{Results from simulations and test rigs demonstrate that the suggested framework is effective at identifying fraudulent users and identifying faulty smart meters.} & \multicolumn{1}{p{14em}|}{This framework is not able detect theft attack that evades the
balance check.} \\

\hline

\multicolumn{1}{|p{0.4em}}{12} &
\multicolumn{1}{|p{13em}|}{Anomaly detection algorithm that analyzes the voltages collected from SM\cite{yen2019effect}} & \multicolumn{1}{p{15em}|}{Based on simulated smart grid model to test the effectiveness of the frequent-and-large smart meter data and compared with three different operating conditions.} & \multicolumn{1}{p{14.5em}|}{Results from simulations and test demonstrate that shorter-duration anomalies can be detected effectively with more frequent data.} & \multicolumn{1}{p{14em}|}{ Lacing of advance decision-making and control system which ensure to facilitate more efficient grid operation and enhance the smartness of the power system.} \\

\hline

\end{tabular}
\end{table*}%

\section{Research Trends and Direction}
The primary goal of this study is to learn about the different existing pertinent approaches for detecting anomalies using a wide range of network packet data sets of the smart grid. However, currently predicting technique of anomalous behavior in SG is mostly based on state-of-the-art Machine Learning (ML) algorithms. Hence, it is therefore necessary to consider the potential data set that is SM data which is a key element in power consumption analysis in SG besides only network or other system data. Particularly for anomaly analysis, SM has emerged as a powerful tool among a range of AMI technologies. Recognizing customer load patterns along with the reason analysis techniques based on SM data is a pivotal step in the research domain in this field. As SM data deals with big data on a daily basis which is a powerful indicator for assessing the grid functionality and operation so one potential approach has been recommended here not only to predict the anomalies but also can be used to predict load patterns and the defense model accordingly. 
To the best of our knowledge, there has not been enough published research work done on the application of these crucial SM data for outage/anomaly identification in SG technology.
After analyzing the gap in the reviewed paper an approach is proposed to identify unusual occurrences or anomalies using the smart meter data at both the lateral and the consumer levels. A generative model for detecting anomalies has been recommended that takes into consideration the information gathered from SM.

\section{Anomaly Detection Approach}

The  anomaly detection approach can be interpreted for multi-level anomaly detection. This model has the ability to efficiently integrate the large-scale data that is routinely collected from SMs at the customers' premises and transform them into actionable real-time insights regarding the anomaly of interest and its severity of it. Fig. 3 shows the essential steps of the recommended machine learning model for energy consumption-based anomaly identification in SG.

The proposed approach is divided into two steps: 1) Method 2) Objective. In the first phase, the overall methodology of this approach has been illustrated ranging from collecting SM data to detecting an anomaly in power consumers' energy usage pattern by the steps of Machine Learning classification which involves feature extraction after the data pre-processing. The primary goal of this model is to recognize the unusual or anomalous load pattern among the customers from the SM power consumption data and based on which the decision-making is performed by the concerned authority to make necessary changes in energy policy. In such a way, working towards the implementation of such a system can be useful to support the concept of ”smarter grids”.

\begin{figure}[h!]
\centering
\includegraphics[scale=0.30]{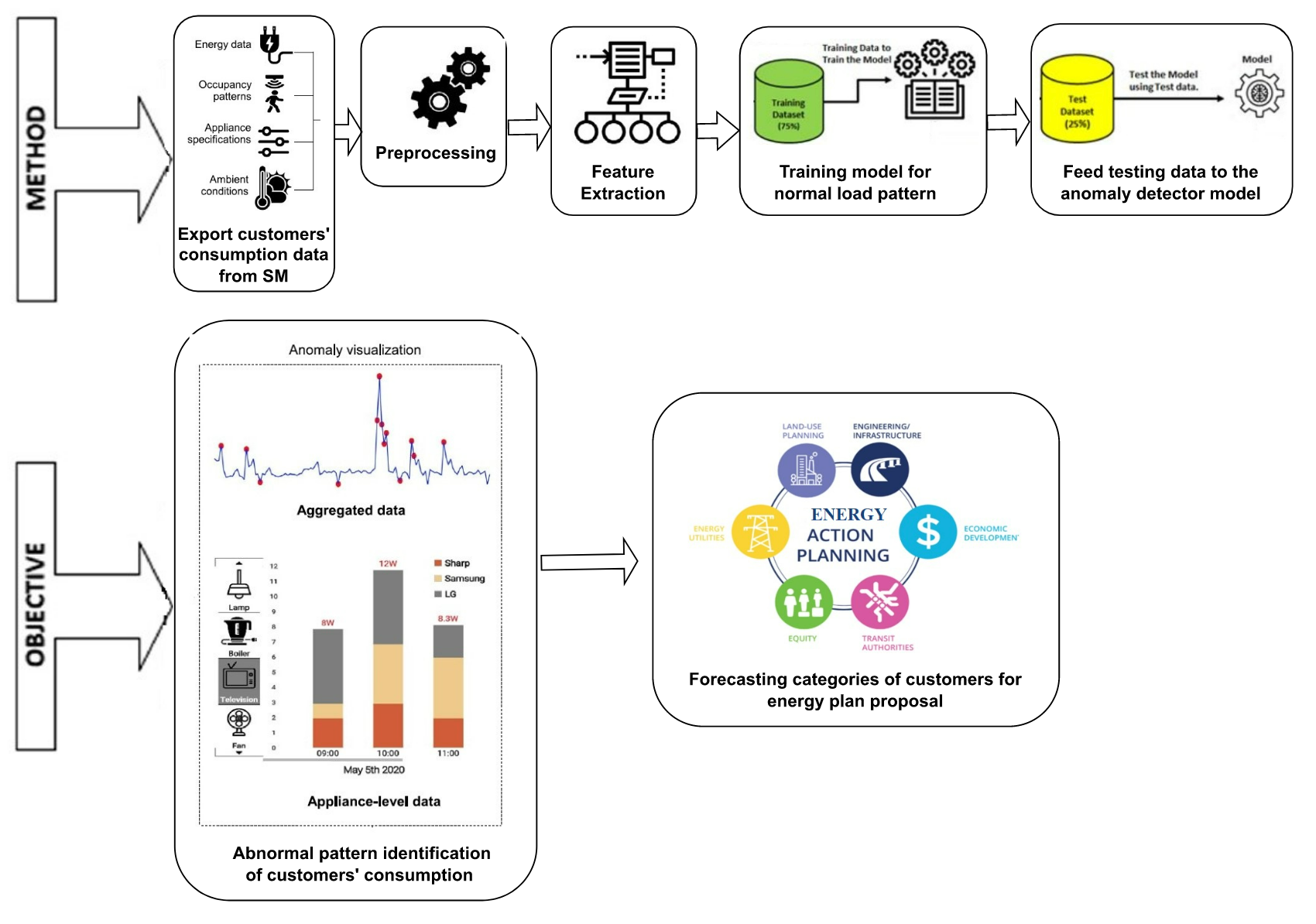}
\centering\caption{Flowchart of main steps used to perform anomaly detection of energy consumption in SG}
\label{Fig. 3}
\end{figure}

In this paper, details of the implementation of the suggested model have not been explored in an experimental way.
Developing the suggested anomaly detection approach has several following common and domain-specific challenges and
limitations, that hinder the development of the best-fit solutions, and limit their implementation widespread.
One of the challenges is the absence of annotated datasets that assists in labeling both normal and abnormal power consumption. Another challenge is to handle the imbalanced dataset that may adversely affect the sub-optimality of the algorithms’ performance. On the other hand, in most cases the distinction between normal and abnormal intake is not well defined, nor are the boundaries between normal and abnormal conduct evident. Additionally, there is a lack of uniform metrics and ground-truth data that may be used to assess the effectiveness of anomaly detection systems.

To address the two aforementioned challenges, for the absence of annotated datasets, generating synthetic data could be considered and on the other hand to deal with large dimensional data, which is a significant difficulty for real-time anomaly detection in SG, the rule-based algorithm such as ”Big Data Analytics in SG” could be introduced which is capable of processing large-size data of diverse categories routinely acquired from SMs. Furthermore, making the distinction between normal and abnormal pattern leveling for the comparison between Anomaly level 1 and Anomaly level 2 could be efficient. Because the genuine stochastic distributions of anomaly levels 1 and 2 are not significantly different from one another(relative to normal vs. anomalous situations), it can be seen that comparing two levels of anomaly is less accurate than comparing normal with anomalous occurrences.

\section{Conclusion}

In this study a systemic and technically-informed survey of
anomaly detection methods in SG have been presented. That is followed by the finding of possible research gaps to develop a new approach to detecting anomalies in the SG systems. Despite the significant research have been made in anomaly detection analysis of SG that is incorporated with the wide spread of network data, the importance of employing SM data has not been explored on a large scale. In order to develop a robust SG with a self-healing nature it is therefore needed to consider the daily collected huge amount of SM data for identifying the anomaly besides other kinds of data. In this context, a new anomaly detection technique is suggested to conclude our study that combines multivariate load data gathered from many SMs to offer insights about anomalies at the customer level. In our upcoming work, as explained and demonstrated in the paper, the platform can be utilized to study the reasoning behind anomalous behavior by the integration of data from many sources (such as substations, feeders, laterals, and customers) and thus it will offer real-time insights for anomaly detection study in SG technology.

\end{document}